\documentclass[sn-mathphys-num]{sn-jnl}%

\usepackage{graphicx}%
\usepackage{multirow}%
\usepackage{amsmath,amssymb,amsfonts}%
\usepackage{amsthm}%
\usepackage{mathrsfs}%
\usepackage[title]{appendix}%
\usepackage{xcolor}%
\usepackage{textcomp}%
\usepackage{manyfoot}%
\usepackage{booktabs}%
\usepackage{algorithm}%
\usepackage{algorithmicx}%
\usepackage{algpseudocode}%
\usepackage{listings}%
\theoremstyle{thmstyleone}%

\theoremstyle{thmstyletwo}%

\theoremstyle{thmstylethree}%

\usepackage{booktabs}                  %
\usepackage{lipsum}                    %
\usepackage{mwe}                       %
\usepackage{nicematrix}

\usepackage{mathptmx}                  %

\usepackage{balance}  

\usepackage{makecell}

\usepackage{arydshln, booktabs}
\usepackage{setspace}
\usepackage{colortbl}
\usepackage{xcolor}
\usepackage{enumitem}

\definecolor{gr1}{rgb}{0.3, 0.3, 0.3}
\definecolor{gr2}{rgb}{0.7, 0.7, 0.7}
\definecolor{gr3}{rgb}{0.95, 0.95, 0.95}

\begin{document}

\title[Article Title]{%
Empirical %
Study on the Representation of 3D Scatterplots as 2D Figures}

\author*[1,2]{\fnm{Philippos} \sur{Papaphilippou}}\email{pp1d24@soton.ac.uk}

\author[2]{\fnm{Lucy} \sur{Hederman}}\email{hederman@tcd.ie}

\affil[1]{\orgdiv{School of Electronics and Computer Science}, \orgname{University of Southampton}, \orgaddress{\street{University Road}, \city{Southampton}, \postcode{SO17 1BJ}, \country{United Kingdom}}}
\affil[2]{\orgdiv{School of Computer Science and Statistics}, \orgname{Trinity College Doublin}, \orgaddress{\street{College Green}, \city{Dublin}, \postcode{D02 PN40}, \country{Ireland}}}

\abstract{3D scatterplots are a well-established plotting technique that can be used to represent data with three or more dimensions. On paper and computer monitors they are essentially two-dimensional projections of the three-dimensional Cartesian coordinate system. %
This transition from the 3D space to two dimensions is not done consistently among scientific software, as there is currently limited quantifiable evidence on the effectiveness of each approach. Notably, the frequent lack of visual cues such as with regard to depth perception is equivalent to a reduction of dimensionality by one. 
Hence, their use in manuscripts is less common or straightforward. In this empirical study, an online survey is conducted within an academic institution to identify and quantify the effectiveness of feature or feature combinations on 3D scatterplots in terms of reading time and accuracy.}


\maketitle

\section{Introduction}\label{sec1}

Effectively visualising and analysing multi-dimensional data is of prime importance 
for a variety of sciences, and is an invaluable tool in data mining, machine learning and statistics. 
With global trends and events such as large-language models \cite{isaev2023scaling} and pandemics \cite{canhoto2022pandemic}, %
intuitively visualising large or complex data %
are %
already concerning a wider demographic.
A visualisation method that is appropriate for multi-dimensional data is 3D scatterplots, as it can translate data for three or more dimensions into the 3D space, but extra care is needed to produce effective and informative plots.

When compared to 2D scatterplots, having an additional dimension in space has the potential to carry dramatically higher amounts of information, but this comes at a cost. %
For instance, a 3D scatterplot can give more complex insights on how each of the dimensions interact with each other. %
However, in order to fit in two-dimensional means such as in scientific manuscripts, they need to be flattened. These essentially reduce to static 2D projections of (pseudo) 3D models. Such projections usually lack realism and visual cues to help with depth perception, for example, and this introduces %
ambiguity in the readings.

This paper explores %
certain visual cues that are hypothesised to significantly help with the readability and effectiveness of 3D scatterplots. %
The explored visual cues include the field of view (FoV), fog, depth-based colour mapping, projections, light paths and surface ticks. These are elaborated in the list below in the form of hypotheses. %
We %
evaluate the effectiveness of these visual cues using random data with a survey of 57 participants from a scientific %
community. The survey measures the time and accuracy of the user measurements to identify impactful visual cue combinations and settings.

\begin{figure}[h!]
\centering
\includegraphics[width=0.80\textwidth, trim=0 0 0 0]{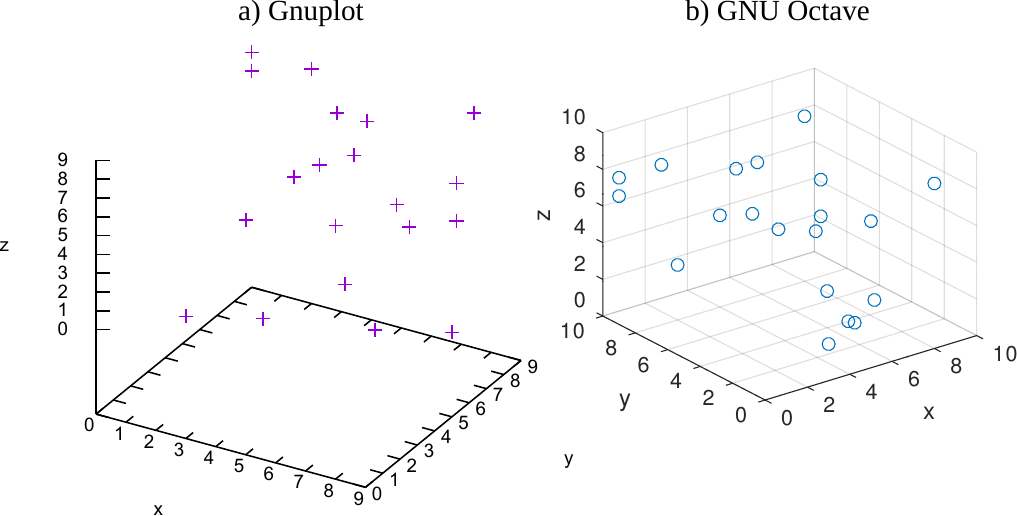}
\caption{Default 3D plots from modern software on an arbitrary dataset}\label{giggnu}
\end{figure}

As a motivation, %
figure \ref{giggnu} shows 3D scatterplots from popular open source plotting software (Gnuplot and GNU Octave) that demonstrate common weaknesses. The default configurations (without additional scripting or code) yield same-sized points, isometric axes, %
and %
absence of additional indicators to track the coordinates %
of the points. %
These example plots use a random distribution dataset as a worse case (as in section \ref{quesamp}), but the discrete nature of the data is easily relatable to existing problems. %
These plots can be considered counterexamples for effectively plotting 3D scatterplots%
, though more elaborate design decisions and software are capable of adding more realism \cite{origin}.

The contribution of this paper is a systematic empirical user study that evaluates a number of visual cues in a quantifiable way in terms of measurement error (accuracy) and time needed to take the measurements (also relating to intuitiveness). %
The visual cues are organised in question types (see section \ref{quesamp}).
The outcomes of this study can be used to increase the effectiveness of 3D scatterplots and plotting software by introducing any missing beneficial features, or enabling them by default. The identified gap in research is the absence of such a quantitative study, %
which can be the reason for software of highly-varying quality in 3D scatterplots, and their relatively limited use on paper. %

\vspace{0.3em}
A series of hypotheses relating to the readability of 3D scatterplots are investigated:
\begin{itemize}%
\item[H1.] An exaggerated field of view (FoV) can be used to improve the readability of 3D scatterplots by improving depth perception.
\item[H2.] Adding fog to a 3D scatterplot can improve the readability of 3D scatterplots by improving depth perception.
\item[H3.] Applying a colour mapping corresponding to the distance from the object to the camera (observer) can improve the readability of 3D scatterplots by improving depth perception.
\item[H4.] Adding projections to the points that correspond to their size and shape can improve their readability by being able to distinguish them more easily on the corresponding grid lines.
\item[H5.] Using semi-transparent surfaces as 2D-equivalents of grid lines on ticks to improve point readability by increasing spatial perception in at least the dimension of the ticks.
\item[H6.] Adding light paths (similar to drop lines) to the points that correspond to their size and shape can improve readability by enhancing the association of projections to points.
\item[H7.] Certain combinations of the above can improve the readability of the points by complementing each other such as with respect to providing visual cues about different dimensions.
\end{itemize}
\vspace{0.3em}

Following is a literature review on the area (section \ref{rw}), the methodology of the survey (section \ref{surv}),  results (section \ref{res}), and finally a discussion (section \ref{disc}) on the hypotheses and limitations, and a conclusion (section \ref{conc}).

\vspace{0.5em}

\section{Related work}\label{rw}
There have been multiple user studies to quantify the effectiveness of certain visualisation techniques. More focusing on interactivity, there are documented use cases where university students appreciated the demonstration of simulations in STEM (science, technology, engineering, and mathematics) lectures. These include a graphical user interface for learning physiology (biology) \cite{hwang2012review} and a computer science visualisation for teaching clustering \cite{fuchs2019educlust}. Both of these studies have a larger corpus of users, but are less quantitative or objective than the presented study, since their performance metrics are rather indirect (relating to satisfaction or learning efficiency).

More related to 3D scatterplots, Yalong et al. have studied the navigation of 3D scatterplots within virtual reality, especially focusing on zooming capabilities \cite{yang2020embodied}. This approach is (indirectly) related to the field of view, as a virtual reality headset is used to facilitate the change of perspectives. %
Sanftmann and Weiskopf%
 conducted a user study on an innovative approach where 2D projection snapshots are frequently introduced while navigating through the points \cite{sanftmann20123d}.  
The participant sample size was only 12, and the main performance metric is navigation success ratio, which is based on binary outcomes per task (true or false). %
This study emphasised the learning rate by conducting longer 1-hour sessions, and produced results based
 on mathematical modelling such as with logistic regression. In contrast, our presented study is on a relatively larger scale, and uses more quantifiable metrics for a different usage %
 of 3D scatterplots as static 2D figures. %
 
There are also research %
works directly targeting the accuracy of scatterplot readability. %
Shovman et al. have used accuracy and time measurements %
to evaluate 3D scatterplots, but this was in regard to how an interactive navigation impacted the readings. An older relevant study showed how an interactive focusing functionality can improve the perception of data \cite{piringer2004interactive}. %
Sanftmann et al. have targeted existing weaknesses of 3D scatterplots to improve %
information transfer \cite{sanftmann2009illuminated} by making heavy use of illumination, aiming at photorealism. 
As with the aforementioned works on 3D scatterplots, these are more focused on interactivity than self-contained static plots, and also use isometric axes. A %
 literature survey has compared a series of sampling methods for the effectiveness of 2D scatterplots \cite{yuan2020evaluation}.
Notably, the majority of the explored techniques rely heavily on spatial localities, such as for medical imaging applications, and could be inappropriate for more discrete datasets. %

Depth perception in humans is a long studied 
subject for both images and plots. Burge et al. study how convexity can affect depth perception, such as by examining luminance and range images based on scenes in the physical world \cite{burge2010natural}. %
Ostnes et al. overview a range of models and techniques for improving depth perception in plots based on real data, such as by using colour maps or through stereoscopy \cite{ostnes2004visualisation}. The effect of the field of view on both time and accuracy has been thoroughly studied for large scale plots in the past \cite{ball2008effects}, though this was with regard to the peripheral vision and the amount of information that fits into computer monitors, rather than to aid depth perception for 3D scatterplots.

\section{Methodology}\label{surv}

\begin{figure}[h!]
\centering
\includegraphics[width=0.7\textwidth, trim=0 0 0 0]{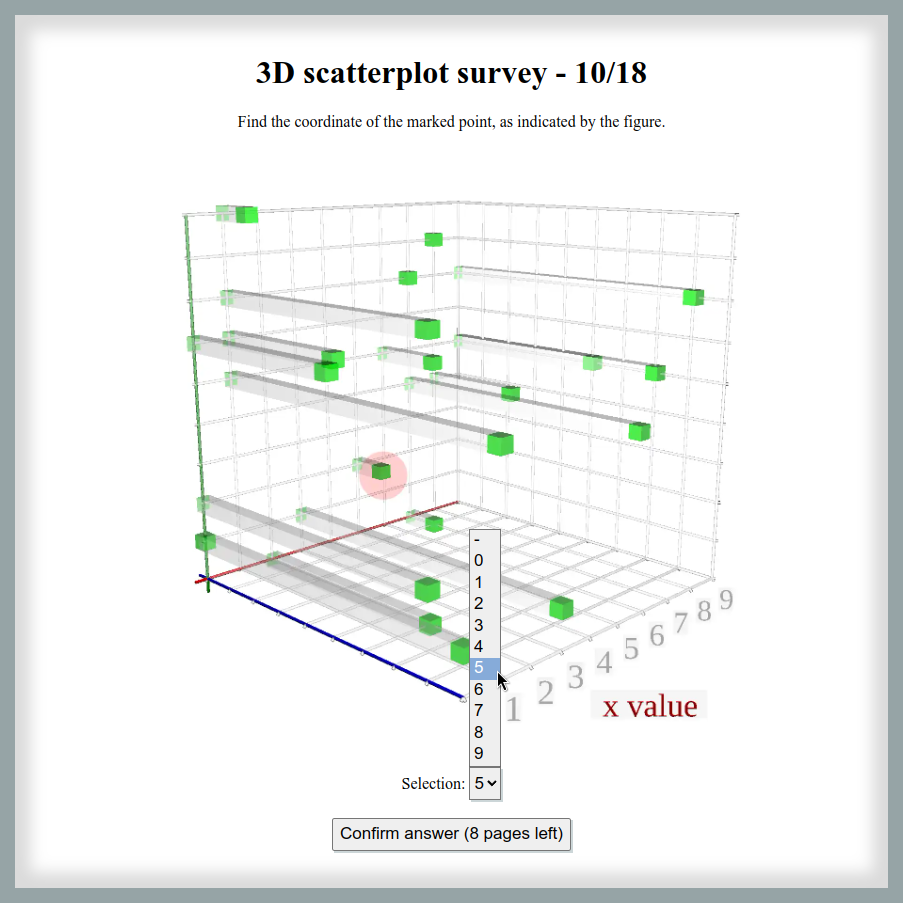}%
\caption{Screenshot of a survey question for the \(x\) value of a point. Note that the questions come in triples, one for each \(x\), \(y\) and \(z\) coordinate. %
}\label{figsc}
\end{figure}

The study is conducted online 
with the aim to identify and quantify the %
effectiveness of different visual cues in 3D Cartesian scatterplots. This is done by measuring the accuracy of  given readings as perceived by participants and the time needed to provide them. The invited participants click on the study link in an email to answer 6 questions of the style ``what is the \([X, Y, Z]\)-coordinate of the marked data point?''. These 6 questions appear on 18 pages, 3 for every one of the axes, to simplify the presented instructions. The number 6 is selected based on the expected time to complete the quiz (see sections \ref{quesamp} and \ref{limit}). The plotted data follow a random uniform distribution per axis, resulting in %
discrete data. By doing so, a worst case is attempted with respect to any data localities in the dataset that may serve as additional visual cues. Figure \ref{figsc} presents an example screenshot from the website during a test session.

\begin{figure*}[h!]
\centering
\includegraphics[width=1\textwidth, trim=0 0 0 0]{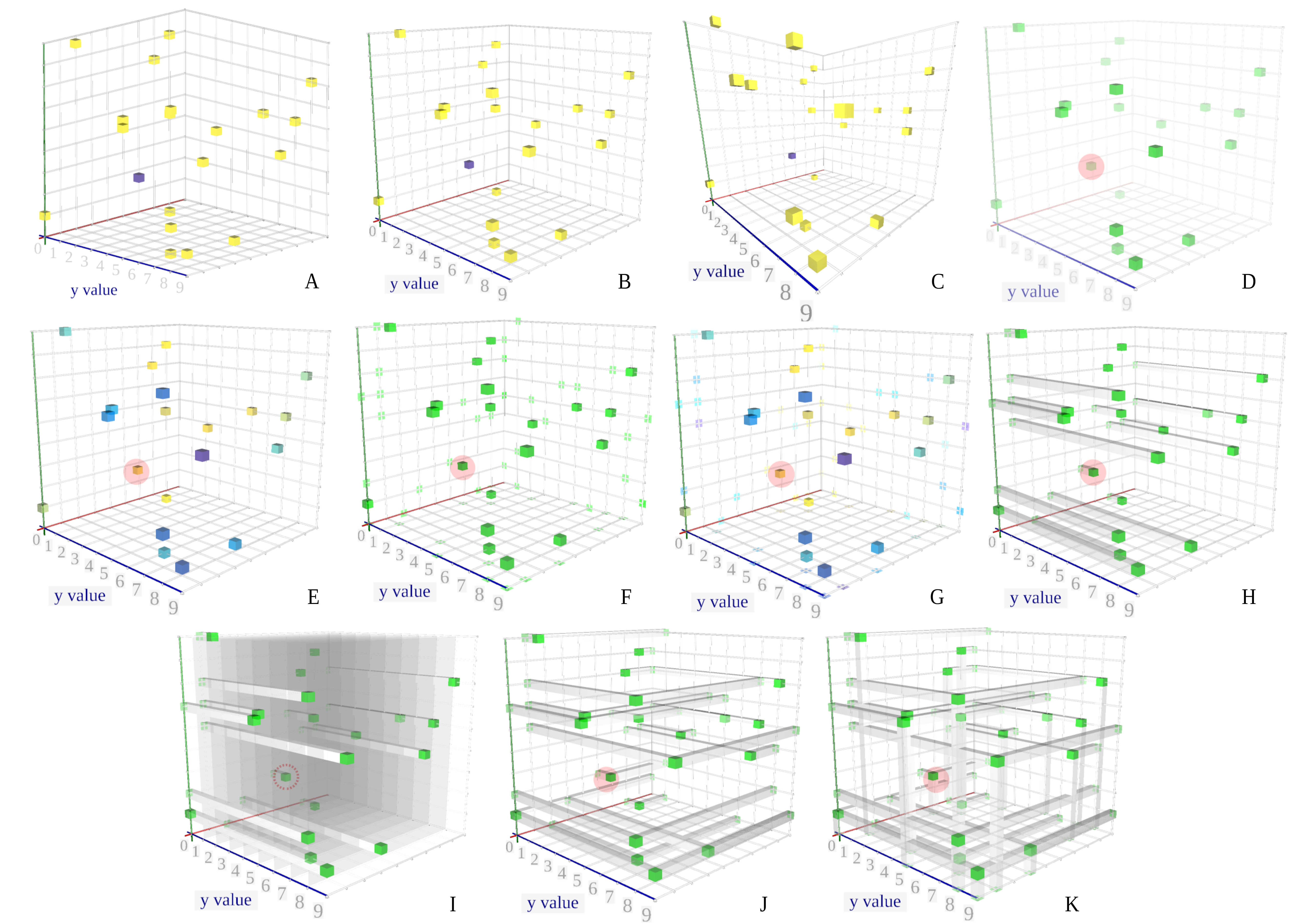}
\caption{Image example per question type, using the same dataset and requesting the \(y\)-coordinate of the same highlighted point. Note that all \(x\), \(y\) and \(z\) coordinates are asked by the survey.}\label{figex}
\end{figure*}

The explored visual cues are with regard to either depth perception or additional indicators that ease the mapping of points to coordinates. For improving depth perception, we explore the \textit{field of view} (FoV), \textit{fog}, \textit{depth-based colour mapping} (the closer an object is to the camera, the darker it appears, as in figure \ref{figex}E).
For easing reading: \textit{projections} (2D ``shadows'' on each pane in the background), \textit{light paths} (3D drop lines, as in figure \ref{figex}H) and \textit{surface ticks} (use of 2D semi-transparent surfaces parallel to the expected ticks on each axis, as in figure \ref{figex}I). %

\subsection{Question sampling}\label{quesamp}

The survey is designed to be informative yet short and inviting, hence there are 11 question types in total, taking up to around 5 minutes to complete. Each question type is a feature combination, and the design space exploration is not exhaustive. %
For instance, the three evaluated values of the field of view (FoV) in questions A, B and C do not multiply the rest of our design space by three, as a default is used as a baseline for %
other questions. %
A future study could utilise the presented results to build new explorations that focus on specific aspects of this study and new combinations of visual cues (see section \ref{limit}). Table \ref{tabq} summarises the question types in terms of visual cue combinations. Figure \ref{figex} shows an example of each question type using the same random dataset.

\begin{table}[h!] 

\caption{Visual cue combinations in the survey}
\label{tabq}
\begin{tabular*}{\textwidth}{@{\extracolsep{\fill}}c} 
\end{tabular*}
\vspace{-1em}
\small
\centering
\setlength{\tabcolsep}{3.5pt}
\setstretch{1.15}

\begin{NiceTabular}{>{\columncolor{gr1!9}}c c >{\columncolor{gr2!9}}c c >{\columncolor{gr2!9}}l c >{\columncolor{gr2!9}}l}[colortbl-like]
\rowcolor{gr1!9}
Question&FoV ($^{\circ}$)&Fog&Colour&Projections&Surface ticks&Light path\\[-0.5em]
\cellcolor{gr3!15} &&\cellcolor{gr3!15}&&\cellcolor{gr3!15}&&\cellcolor{gr3!15}\\
A&3&&mark$^1$&&\\
B&30&&mark$^1$&&\\
C&75&&mark$^1$&&\\
D&30&yes&&&\\
E&30&&depth$^2$&&&\\
F&30&&&\(xy,xz,yz\)&&\\
G&30&&depth$^2$&\(xy,xz,yz\)&&\\
H&30&&&\(xz\)&&\(xz\)\\
I&30&&&\(xz\)&\(y\)&\(xz\)\\
J&30&&&\(xz,yz\)&&\(xz,yz\)\\
K&30&&&\(xz,yz,xy\)&&\(xz,yz,xy\)\\
\end{NiceTabular}\\
\vspace{0.8em}
\small
$^1$ The point colouring here is  used to identify the point in question.\\
$^2$ The point colouring here is used as a visual cue for depth perception.
\setstretch{1}

\end{table}

The %
rationale behind this selection of questions is to be able to address the hypotheses as introduced in section \ref{sec1}.  Hypothesis H1 on FoV is investigated by the combination of question types (A, B, C), hypothesis H2 on fog by the combination (B, D), hypothesis H3 on depth-based colour-mapping by questions (B, E) and hypothesis H5 on surface ticks by questions (H, I). Hypothesis H4 on projections is mainly explored by (B, F), and H6 on light paths is explored incrementally with (B, H, J, K) corresponding to 0, 1, 2 and 3 light paths. Hypothesis H7 on combinations is partly addressed by questions with overlapping features such as G that enables colour-mapping and projections, or by the incremental aspect of some of the question combinations like (B, H, J, K) for light paths, as well as their assumed coexistence with their corresponding projections. See section \ref{pairw} for the final discussion on the hypotheses. %

There are 6 random datasets, each containing 10 data points with their \(x\),\(y\) and \(z\) coordinates %
 selected at random. The random values are the output of a function drawing samples from a uniform distribution of integers within the range \([0,9]\). One of the 10 points in each dataset is randomly selected as the point in question, after which the images are marked accordingly (e.g. using a red circle) for the pages to direct the question on the selected points. The reason for choosing %
10 possible answers per axis is to get a certain variation in the responses while still being meaningful to provide a correctness score at the end. The correctness score is not used in the study as a performance metric, but serves as a reward. %

Since there are 11 possible question types and 6 random datasets applied to each question, there are a total of 66 unique questions. In order to find the total number of unique pages that can be displayed by the website, this number is multiplied by 3, as there are separate images for each of the pages requesting the \(x\),\(y\) and \(z\) coordinates (totalling 198). Once a question is selected, its 3 pages appear consecutively to the user, so that a 3D distance is used as a performance metric. Being consistent with regard to the dataset for the same question is also useful for the user friendliness of the quiz, as the user has comparatively more time to reflect on the same dataset. 

This number of possible question type combinations is \({11 \choose 6}=462\), since each user gets a random subset of the 11 question types of size 6. Every question appears with a unique dataset to the others, and a random permutation of the datasets increases the uniqueness of each quiz instance with respect to the dataset order. After considering the \(6!=720\) dataset permutations, we conclude that there are \(720\times462 = 332,640\) possible quiz instances, with varying amounts of similarity. The most important feature here is the uniqueness of the datasets between the questions of the same instance, to avoid unintentional knowledge transfer %
between different questions. Finally, considering multiple datasets per question is crucial for disassociating any biases related to the difficulty of the dataset (elaborated in section \ref{res}).%

\subsection{Software infrastructure}

The quiz is hosted on a virtual instance on Amazon AWS running Debian Linux. It is served as a publicly-available website using the Apache HTTP Server (%
though the link is only circulated in our academic department). It is coded as a Python script, which provides the required HTML code (HyperText Markup Language) in the standard output. %
The script is arranged to be recognised by the CGI (Common Gateway Interface) module of Apache, which executes it every time the %
link is accessed. 

The %
script is written from scratch and facilitates the dynamic aspect of the quiz as a website. As opposed to static websites, the CGI approach is used to be able to discreetly process the information, select the questions, save the results, and announce two random winners of a 25-pound Amazon voucher at the end. The output is stored locally on the server for further processing by the authors. For a larger-scale study, it would be appropriate to use a more well-rounded solution such as with a separate front-end, back-end and database infrastructure for added security and scalability. %

Alongside the script, the server hosts the 198 images in a subfolder. The script injects those images into the HTML code using the base64 format. This is done in order to minimise the possibility of giving more information based on the image filename, or enabling the crawling of all the images. These images have been generated in a semi-automated way, as the current software is GUI-focused%
. Another Python script has been used to generate the 6 random datasets, and then the GUI was used (incrementally) to export the images. The marking of the random points is done manually using GIMP relatively quickly by opening all images in tabs at once. Finally, the images are overwritten in the \texttt{.png} format (Portable Network Graphics), though a separate bash script compresses them into the more modern and space-efficient \texttt{.webp} format (WebP). The resulting images have a resolution of 1700 by 1275, and consume up to 60 KB each, which also eases their serving through HTML injection.

Juniper \cite{juniper} is the software that is used to generate the 3D scatterplots. It is developed in-house and is open-source \cite{jun}. It is written in Java and uses JavaFX and FXyz for the 3D functionality. The plots and their labels exist as 3D models in their entirety, which enables the support of all of the aforementioned visual cues.

\subsection{Participation}

Participation %
is voluntary, %
and an invitation email has been circulated in certain mailing lists. The invitation email was sent to academic staff and students of the %
School of Computer Science and Statistics (SCSS) at Trinity College Dublin, Ireland. 
Although there was %
no special criterion for any particular engagement with plotting methods (such as visualisation research), a certain degree of plot comprehension by a scientific community is expected. The final sample size for the current study is 57, totalling 31.1 data points per question type on average. %

Before conducting the study, an ethics application was approved by the research ethics committee of the School.
 The resulting dataset is completely anonymous, as it consists only of the coordinates of random points and the time that was needed to find them. A ``user agent'' string per session with browser information is temporarily kept on the server for diagnosing technical issues. %
Before starting each quiz session, there is a consent checkbox that needs to be selected for a ``Start'' button to be activated. %
The hyper-linked consent page is also sent as a separate document inside the %
invitation email. 

\section{Results}\label{res}

\begin{figure}[h!]
\centering
\includegraphics[width=0.7\textwidth, trim=0 0 0 0]{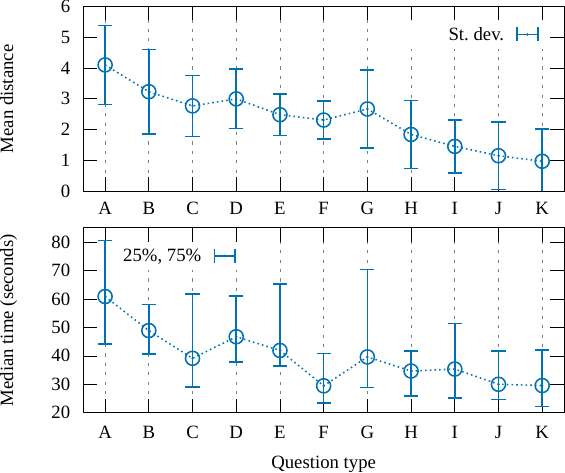}
\caption{Mean distance and median time taken for each question type}\label{figres}
\end{figure}

The unprocessed set of results is presented in figure \ref{figres}. The top plot of the figure is the mean distance in the 3D space of the %
perceived coordinates of the point to the actual data point. Also shown are the standard deviations of the corresponding mean values. As described, the \(x\), \(y\) and \(z\) coordinates of the marked points are requested from the participant in three consecutive pages, one for every coordinate, and this information is enough for a single distance measurement per question type per user. 

The lower plot of figure \ref{figres} shows the median time a user needed to answer the corresponding questions. The time corresponds to the sum of the time for all 3 coordinate responses. Although time is a more noisy variable than the explicitly requested distance (hence showing the median with 25th and 75th percentiles), there is still a certain amount of consistency and value in the findings. %

\subsection{%
Basic insights}

As a first observation, %
types J and K (2 and 3-axis light paths) provide the most accuracy, while the type A (FoV near 0$^{\circ}$) and B (FoV of 30$^{\circ}$) seem to be the most ineffective. J and K perform similarly, and this is expected because even with 2 light paths (J), there is direct information for every coordinate. This is because each path points 
to a projection %
on one of the \(xy, xz\) or \(yz\) surfaces, and any 2-selection assists the readings of all axes. At a first glance, F (projections) also has promising timing behaviour, though this is better explored in a later discussion. With respect to A, the low performance validates the issue as presented in the motivation of this paper, which is the flaw of common plotting software using isometric axes for 3D plots. D (fog), however, even with an additional cue from B (with FoV=30$^{\circ}$), performs similarly to B. The latter could also be attributed to the lack of training or familiarity with the use of fog %
to measure distance in a small space.

Another interesting observation is about the timing of A, B and C, as the higher the FoV the lower the time. %
Although C is generally better than B, there are some outliers with respect to the time, and this could be attributed to the fact that a FoV of 30$^{\circ}$ (B) is more natural to the human eye \cite{cha1992mobility} (e.g. corresponding to a nearby computer screen \cite{van2007effect}). There are some other timing outliers, as with E and G (depth-based colour mapping), probably because they are rather unconventional and require %
some extra time to process. %
Colour mapping is more conventionally mapped to a single axis (height) rather than the depth. %

\subsection{%
Normalisation}\label{norm}
In order to remove potential biases, such as with a variation of the user skill or the difficulty of the random dataset, we apply normalisation at different levels. The goal is to increase the reliability of the time/distance comparison especially for more subtle observations. %

The first level of normalisation is about the user competency. The distance measurements are now normalised by the average distance per user. In other words, every distance value per user (and question type) is divided by the mean distance of all distances for that user. 
Equivalently, this is done for the time as well. The denominator in this case is the median instead of mean time, since the time is relatively more erratic. %
In order to calculate this median, the time of a question response represents the median response time between the results for the three axes. This can be seen as a more aggressive %
filtering step to remove timing outliers, such as when a participant is %
distracted. 

Another normalisation step is about the difficulty of the dataset. For instance, in the isometric view case, the further out a data point appears in the view box, the less the ambiguity is in regard to the distance from the observer (depth). Still, each visual cue can have different strengths and weaknesses corresponding to each random dataset. %

\begin{figure}[h!]
\centering
\includegraphics[width=0.75\textwidth, trim=0 0 0 0]{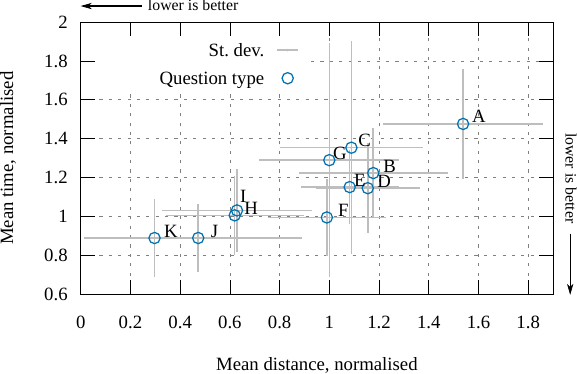}
\caption{Survey results after applying normalisation %
}\label{fignorm}
\vspace{-0.5em}
\end{figure}

Figure \ref{fignorm} presents the normalised results of this exploration. It combines the two performance metrics into a single scatterplot for an easy correlation between them by eye. %
As shown in the plot, the results are fairly conclusive, as the Pareto front is just two points; K (3 light paths and projections) and J (2 light paths and projections).  
Similarly, the most commonly used A is the worst performer in both distance and time by essentially missing an entire dimension.
As for the observations made with non-normalised results, most still hold with normalisation. F (projections) stops being competitive, and there are more visible trade-offs in the center points (outside the Pareto front). This shows the impact of normalisations, but also overall verifies the well-roundness of the results %
especially on the sample size and questions. A more subtle finding is about the pairs that only differ in time. In the first instance, F and G (non-coloured and depth-coloured projections respectively) perform similarly in terms of accuracy, but the colouring could have caused some delay from being an unfamiliar approach. 
In the other instance, C and E are the respective figures without the projections, but here the colour has reduced the time required to do the reading. Overall, there are clear winners, but for points with trade-offs, the  unfamiliarity or information crowding could be %
a notable factor for time required to make the readings. %

\section{Discussion}\label{disc}

This %
section overviews the key findings from the survey with the help of pairwise comparisons, while elaborating on potential limitations and future work on the methodology. %

\subsection{Key survey insights}\label{pairw}

The main observation from the survey is that the most efficient technique is the light paths with projections (types J and K, with 2 and 3 light paths respectively). These correspond to the hypotheses H4 on projections, H6 on light paths, and H7 on combinations, as introduced in section \ref{sec1}, and are all confirmed.
Another important finding relates to the FoV, as the popularly-used isometric axes gave unacceptable performance, while the higher FoV value (75$^{\circ}$) gave the most promising results among the types comparing FoV values. Therefore, H1 on exaggerated FoV is mostly supported, even though the reduced variation when having a moderate FoV is notable. Other less-conventional %
 features such as fog, depth-based point-colouring and surface ticks have not caused a substantial improvement as additional visual cues%
, but this could also partly be attributed to the lack of training. The latter three correspond to hypotheses H2 on fog, H3 on colour-mapping and H5 on surface ticks, which are not supported in this exploration.

\begin{figure}[h!]
\centering
\includegraphics[width=0.80\textwidth, trim=0 0 0 -2]{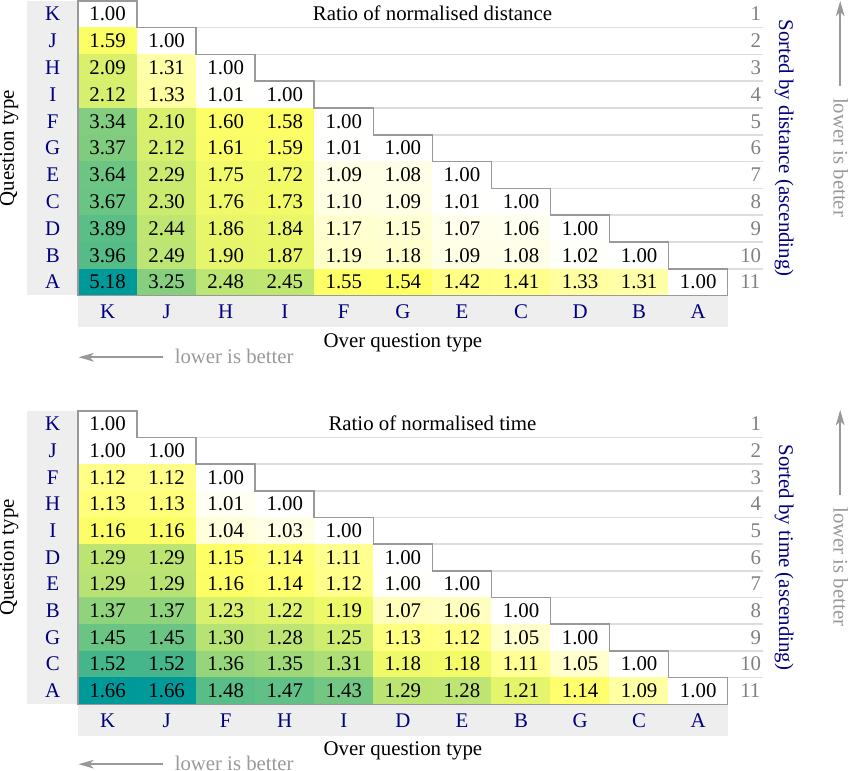}
\caption{Comparing the performance of question types in pairs. %
}\label{comp}
\end{figure}

The colour-mapped matrices of figure \ref{comp} can be used to make direct comparisons between question types. The question types are also sorted according to the corresponding performance metric (distance at the top, time at the bottom), also denoting a ranking. The two rankings are relatively similar permutations of the types, with G (colour-mapped projections) making the most steps in the ranking, from 6th for distance to 9th for time. If we select the best performing (K, three projections/light paths), it can be seen that it provides 5.18x lower distance and 1.66x lower time than the frequently-used isometric plot without additional cues (A). Additionally, incremental changes can be compared directly including the FoV variations, as with 75$^{\circ}$ (C) over isometric (A) that reduces error by 1.41x and time by 1.09x. It is also interesting to observe areas where %
similar shades of green appear clustered, such as with K, J, H and I which perform the best in terms of distance, as they include light paths that point to at least one pane (2-axis combination) for guidance. %

\subsection{Limitations}\label{limit}

A %
limitation in the study %
is that there was no training of the participants, or a monitoring process to ensure that they get familiar with the plot features, and have the same interfacing experience. Another consideration is with colour blindness, and its potential impact on the %
readings, which is not studied here. %
The plotting software uses the viridis colour palette, which is expected to help with distinguishing between colours by colour blind people \cite{rocchini2022scientific}. %
Nevertheless, %
the winning question types (J and K) feature no colour-mapping. 

Some simplifying assumptions have been made to reach a low number of questions, such that the surface ticks are  appropriate to be used on only one axis, complementing a cue on the other two dimensions. For instance, in plot I of figure \ref{figex}, the point projections are on the \(xz\) pane, while surface ticks are on the \(y\) axis. Additionally, the design space has been simplified by ignoring any other FoV values than 3$^{\circ}$, 30$^{\circ}$ and 75$^{\circ}$, a camera tilt and pan angles other than 45$^{\circ}$ and 135$^{\circ}$ respectively, and any other combination of the explored features. The identified helpful features could be combined to further improve the reading performance.

Finally, the study results are partly tied to the nature of the dataset. 3D scatterplots are known to be prone to overplotting, when there are high numbers of point intersections \cite{ball2008effects}. This is important to keep in mind for large datasets, where certain visual cues might become more effective or even contribute to the crowding and occlusion in the plot. Similarly, the study assumes the worst case in terms of spatial localities within the dataset by using a uniform random distribution. This is done to increase the generality of the results%
, but for spatial data, other visual cues are known to improve 3D scatterplot readability, as with illumination \cite{sanftmann2009illuminated}.

\section{Conclusions}\label{conc}
 {
This study shows that certain visual cues can dramatically affect the effectiveness of 3D scatterplots. An online survey requested 57 participants from a scientific community to provide readings from marked 3D points. When compared %
to a conventional plot with isometric axes, the best candidate (wider field of view, 3 light paths and projections) reduces the read error (distance) for discrete data by 5.2 times on average, while also reducing the reading time by 1.7 times. Simpler adjustments also resulted in conclusive performance results, such as with the exaggerated field of view of 75$^{\circ}$ which reduces the mean distance and time taken by 1.4 and 1.1 times respectively. %
A series of measures has been applied before and after the data collection to ensure the impartiality and appropriateness of the results. These include the uniqueness between survey questions and instances, and the normalisation steps on the user competency and the difficulty per random dataset. {\Large \color{white}IO}
}

\setstretch{1}

\balance

\bibliography{bibl}

@article{canhoto2022pandemic,
  title={The pandemic-induced personal data explosion},
  author={Canhoto, Ana Isabel and Brough, Aaron R},
  journal={Social Marketing Quarterly},
  volume={28},
  number={1},
  pages={78--86},
  year={2022},
  publisher={SAGE Publications Sage CA: Los Angeles, CA}
}

@inproceedings{isaev2023scaling,
  title={Scaling infrastructure to support multi-trillion parameter LLM training},
  author={Isaev, Mikhail and McDonald, Nic and Vuduc, Richard},
  booktitle={Architecture and System Support for Transformer Models (ASSYST@ ISCA 2023)},
  year={2023}
}

@article{rocchini2022scientific,
  title={Scientific maps should reach everyone: a straightforward approach to let colour blind people visualise spatial patterns},
  author={Rocchini, Duccio and Nowosad, Jakub and D'Introno, Rossella and Chieffallo, Ludovico and Bacaro, Giovanni and Gatti, Roberto Cazzolla and Foody, Giles M and Furrer, Reinhard and G{\'a}bor, Luk{\'a}{\v{s}} and Lovei, Gabor L},
  year={2022},
  publisher={EcoEvoRxiv}
}

@article{burge2010natural,
  title={Natural-scene statistics predict how the figure--ground cue of convexity affects human depth perception},
  author={Burge, Johannes and Fowlkes, Charless C and Banks, Martin S},
  journal={Journal of Neuroscience},
  volume={30},
  number={21},
  pages={7269--7280},
  year={2010},
  publisher={Soc Neuroscience}
}

@article{ostnes2004visualisation,
  title={Visualisation techniques: An overview-part 2},
  author={Ostnes, Runar and Abbott, VJ and Lavender, Samantha},
  journal={Hydrographic Journal},
  pages={3--10},
  year={2004},
  publisher={THE HYDROGRAPHIC SOCIETY}
}

@article{sanftmann20123d,
  title={{3D scatterplot navigation}},
  author={Sanftmann, Harald and Weiskopf, Daniel},
  journal={IEEE Transactions on Visualization and Computer Graphics},
  volume={18},
  number={11},
  pages={1969--1978},
  year={2012},
  publisher={IEEE}
}

@inproceedings{piringer2004interactive,
  title={{Interactive focus+ context visualization with linked 2D/3D scatterplots}},
  author={Piringer, Harald and Kosara, Robert and Hauser, Helwig},
  booktitle={2nd Int. Conf. on Coordinated and Multiple Views in Exploratory Visualization},
  pages={49--60},
  year={2004},
  organization={IEEE}
}

@inproceedings{sanftmann2009illuminated,
  title={{Illuminated 3D scatterplots}},
  author={Sanftmann, Harald and Weiskopf, Daniel},
  booktitle={Computer Graphics Forum},
  volume={28},
  number={3},
  pages={751--758},
  year={2009},
  organization={Wiley Online Library}
}

@article{hwang2012review,
  title={Review of use of animation as a supplementary learning material of physiology content in four academic years},
  author={Hwang, Isabel and Tam, Michael and Lam, Shun Leung and Lam, Paul},
  journal={Electronic Journal of E-learning},
  volume={10},
  number={4},
  pages={pp368--377},
  year={2012}
}

@article{fuchs2019educlust,
  title={EduClust: A visualization application for teaching clustering algorithms},
  author={Fuchs, Johannes and Isenberg, Petra and Bezerianos, Anastasia and Miller, Matthias and Keim, Daniel A},
  year={2019}
}

@article{yang2020embodied,
  title={Embodied navigation in immersive abstract data visualization: Is overview+ detail or zooming better for 3d scatterplots?},
  author={Yang, Yalong and Cordeil, Maxime and Beyer, Johanna and Dwyer, Tim and Marriott, Kim and Pfister, Hanspeter},
  journal={IEEE Transactions on Visualization and Computer Graphics},
  volume={27},
  number={2},
  pages={1214--1224},
  year={2020},
  publisher={IEEE}
}

@article{yuan2020evaluation,
  title={Evaluation of sampling methods for scatterplots},
  author={Yuan, Jun and Xiang, Shouxing and Xia, Jiazhi and Yu, Lingyun and Liu, Shixia},
  journal={IEEE Transactions on Visualization and Computer Graphics},
  volume={27},
  number={2},
  pages={1720--1730},
  year={2020},
  publisher={IEEE}
}

@misc{jun,
author={Papaphilippou, Philippos},
  title={{Juniper: a JavaFX-based Plotting Framework for Effective Multi-Dimensional 3D Scatterplots (Code Repository)}},
  url={https://github.com/pphilippos/juniper},
  note={{[Accessed 12/2025]}},
  year={2025}
}

@article{juniper,
title = {{Juniper: A 3D plotting framework for effective multi-dimensional scatterplots}},
journal = {SoftwareX},
volume = {31},
pages = {102203},
year = {2025},
issn = {2352-7110},
url = {https://www.sciencedirect.com/science/article/pii/S2352711025001700},
author = {Philippos Papaphilippou}
}

@inproceedings{ball2008effects,
  title={The effects of peripheral vision and physical navigation on large scale visualization},
  author={Ball, Robert and North, Chris},
  booktitle={Proceedings of graphics interface 2008},
  pages={9--16},
  year={2008}
}

@article{origin,
author={{OriginLab Corp.}},
title={Origin: Data analysis and graphing software},
journal={{originlab.com [Accessed 08/2024, Tested version 10.1.5.132]}}
}

@article{cha1992mobility,
  title={Mobility performance with a pixelized vision system},
  author={Cha, Kichul and Horch, Kenneth W and Normann, Richard A},
  journal={Vision research},
  volume={32},
  number={7},
  pages={1367--1372},
  year={1992},
  publisher={Elsevier}
}

@article{van2007effect,
  title={The effect of cybersickness on the affective appraisal of virtual environments},
  author={van der Spek, ED},
  year={2007},
  journal={Master’s thesis, Universiteit Utrecht}
}

\end{document}